\documentclass[aps,pra,twocolumn,showpacs]{revtex4}
\usepackage{graphicx}
\begin{document}
\title{Simulating a single qubit channel using a mixed
state environment}
\author{Geetu Narang}
\email{g_n29@yahoo.com}
\affiliation{Department of Physics,
Guru Nanak Dev University, Amritsar 143005, India}
\author{Arvind}
\email{arvind@quantumphys.org}
\affiliation{Department of Physics,
Indian Institute of Madras, Chennai 600036, India}
\begin{abstract}
We analyze the class of single qubit channels with
the environment modeled by a one-qubit mixed
state.  The set of affine transformations for this
class of channels is computed analytically,
employing the  canonical form for the two-qubit
unitary operator.  We demonstrate that,
$\frac{3}{8}$ of the generalized depolarizing
channels  can be simulated by the one-qubit mixed
state environment by explicitly obtaining the
shape of the volume occupied by this class of
channels within the tetrahedron representing the
generalized depolarizing channels.  Further,
as a special case, we show that the
two-Pauli Channel cannot be simulated by a one-qubit
mixed state environment.
\end{abstract} 
\maketitle
\section{Introduction} 
The storage and transmission of quantum states in
a decohering environment is unavoidable in quantum
computation and quantum communication
~\cite{palma-roy-1996,zurek-rmp-2003,nielson-book-2002}.
The conceptual device of a quantum channel is very
useful in addressing the issues of such a
transmission~\cite{schumaker-pra-1996} and to
analyze decoherence related questions in
quantum cryptography~\cite{gsain-rmp-2002} and
quantum teleportation~\cite{bose-prl-2001,yeo-pra-2003}.

A general quantum operation on an $n$-dimensional
quantum system is a completely positive map. Such
a map defines a quantum channel for the given
system and can be formally represented by its
operator sum
representation~\cite{ecgs-pra-1961,choi-alg-1975,kraus-book-1983}.
One can also look at this general quantum
evolution using affine transformations for the
channel~\cite{jordan-pra-2005}.  In an explicit
model for this evolution, the system is considered
as a part of a larger closed system undergoing
unitary evolution. The part of this larger system
that we are not interested in can be thought of as
the environment, which when traced over, gives us
the dynamics of the effective sub-system.  The
environment can in general be very large. However,
it turns out that in order to achieve the most
general evolution for a system with an
$n$-dimensional Hilbert space, we need an
environment which is of dimensions $n^2$ (if we
allow the most general unitary evolution of the
total system and assume that the initial state of
the environment is
pure)~\cite{schumaker-pra-1996}.  Therefore, to
simulate the most general evolution of a single
qubit, at least a four-dimensional (two-qubit)
environment is required. However if we allow the
initial state of the environment to be a mixed
state, there is a possibility of achieving such a
simulation by employing a smaller environment.
This reduction in the dimension of the Hilbert
space is desirable for performing actual
simulations of open quantum
systems~\cite{terhal-pra-1999}.  Along these
lines, an argument based on  counting the number
of independent parameters suggests that a
one-qubit mixed state environment might be
sufficient to simulate the most general quantum
evolution of a single
qubit~\cite{lloyd-science-1996}.  Further
investigations in this direction have revealed
that there are counter-examples to the above
conjecture, and there are single qubit channels
which cannot be simulated by one qubit in the
environment~\cite{terhal-pra-1999,zalka-jmp-2002,becon-pra-2001}.

We investigate this question in detail via a
different route  and compute the set of affine
transformations for this class of channels
analytically, employing the canonical form for
two-qubit unitary operators~\cite{kraus-pra-2001}.
Restricting ourselves to generalized depolarizing
channels, we show that a sizable volume (namely
$\frac{5}{8}$) of these channels cannot be
simulated by a one-qubit mixed state environment.
Further, we show that the counter example of the
two-Pauli channel found by Terhal et. al. 
is a special case of our
results.

The material in this paper is arranged as follows:
in Section~\ref{one-qubit-channel}, we explain how
a one-qubit mixed state environment can be
considered for the simulation of a single qubit
channel and  the significance of modeling the
channel using such an environment.  We consider
the two-qubit unitary required for the simulation
of such a channel and discuss how this class of
channels could be a candidate that could occupy a
sizable volume in the space of  single-qubit
channels.  At the end of this section, the
expression for the affine transformation for a
general single qubit channel modeled using a
one-qubit mixed state environment is obtained.  In
section~\ref{depolarizing-channel} we take up the
special case of generalized depolarizing channels.
The affine transformation for this case is
obtained by setting the shift of the Bloch sphere
origin to zero in the general expression for
affine transformation developed in
Section~\ref{one-qubit-channel}.  These channels
are classified by computing  their singular
values. From the structure of these singular
values and the analysis of all the possible cases
the volume occupied by the generalized
depolarizing channels simulated by a one-qubit
mixed state environment is computed and compared
to the total volume of  generalized depolarizing
channels. The section concludes with a discussion
of the two-Pauli Channel considered as a special
case of generalized depolarizing channels and it
is shown that it cannot be simulated by  a
one-qubit mixed state environment.
Section~\ref{conclusions} contains some concluding
remarks.
\section{One-qubit channels with a one-qubit mixed state as 
the environment}
\label{one-qubit-channel}
\begin{figure}
\includegraphics[scale=1,angle=0]{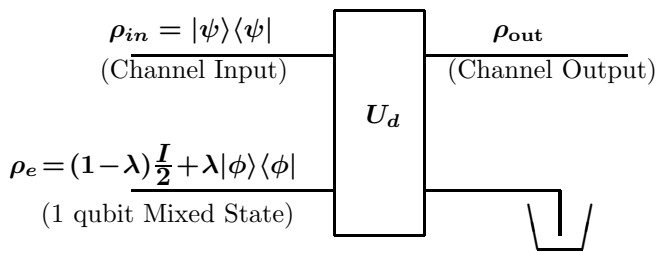}
\caption{\label{channel}
Implementation of a one-qubit channel using a
single qubit mixed state environment. The system
is taken to be in a pure initial state $\vert
\psi\rangle$ while the environment is in a mixed
state $\rho_e$. The interaction takes place via
the two-qubit unitary $U_d$ and the environment
qubit is discarded after the interaction.}
\end{figure}

In the model for the one-qubit channel that we
consider, we allow the qubit of interest to
interact with only one environment qubit. The
interaction unitary is allowed to be the most
general two-qubit unitary and the environment
qubit is a general mixed state. This channel is
schematically depicted in Figure~(\ref{channel}).

For the purpose of characterizing the channel, we
consider the system qubit to be in a general pure
state (a point on the Bloch sphere) given by
\begin{equation} 
\vert
\psi\rangle=\cos{\frac{\theta}{2}}\vert0\rangle +
e^{-i\phi} \sin{\frac{\theta}{2}}\vert 1\rangle
\label{ch13_t} 
\end{equation}
where $\theta$ varies from 0 to $\pi$ and $\phi$
varies from 0 to 2$\pi$.

The environment qubit is assumed to be in a general mixed state
for which we choose a particular parameterization
as follows:
\begin{equation}
\rho_e = (1-\lambda)\, \frac{I}{2} + \lambda\,\vert
\phi\rangle\langle\phi\vert
\label{ch14_t}
\end{equation}
where 
$\frac{I}{2}$ corresponds to a maximally mixed state 
and  $\vert \phi\rangle$ corresponds to a general pure state 
given by
\begin{equation}
\vert \phi\rangle=\cos{\frac{\xi}{2}}\vert0\rangle
 + e^{-i\eta} \sin{\frac{\xi}{2}}\vert 1\rangle
\label{ch15_t}
\end{equation}
where $\xi$ varies from $0$ to $\pi$ and $\eta$ varies from
$0$ to
$2 \pi$.  Therefore, the  environment state can be thought of as a
mixture of a completely mixed state and a pure state.  The
parameter $\lambda $ indicates the degree to which the state is
mixed. 
By varying $\lambda$ from $1$ to $0$, we can go
from a pure to a maximally mixed state of  the environment.

\begin{figure}
\includegraphics[scale=1,angle=0]{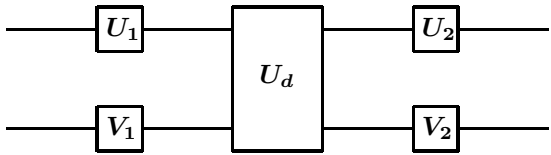}
\caption{\label{u4-fig}
Schematic Diagram for a two-qubit unitary.
The local unitaries $U_1, V_1, U_2, V_2$ together with $U_d$,
which is diagonal in the special Bell basis, 
provide a decomposition of a general $SU(4)$
transformation.}
\end{figure}

The interaction that takes place between the system qubit and the
environment qubit corresponds to a general unitary operator in
the four-dimensional (two-qubit) Hilbert space. This most general
unitary which is an $SU(4)$ transformation has fifteen independent
parameters. However, it has been shown 
that an arbitrary  two-qubit unitary can be decomposed into local
unitaries on each qubit sandwiching a non-trivial
interaction unitary. This interaction unitary $U_d$ belongs
to a three-parameter family of 
transformations~\cite{kraus-pra-2001}.
\begin{eqnarray}
&
U = (U_1 \otimes V_1) \,U_d\, (U_2\otimes V_2) \nonumber\\
&U \in SU(4); \,{\rm and}\, U_1,U_2,V_1,V_2 \in SU(2)
\label{ch5_t}
\end{eqnarray}
This is pictorially shown in
Figure~\ref{u4-fig}.

This three-parameter family of
interaction unitaries 
has the power to entangle or
unentangle the qubits involved. 
This family finds a simple
diagonal representation in the special Bell basis, namely
\begin{equation}
U^{\rm Bell}_d=\sum e^{-i\lambda_i}\vert \psi_i\rangle \langle
\psi_i\vert.
\label{ch6_t}
\end{equation}
where $\vert \psi_j$ for $j=1,2,3,4$ are the special Bell
basis vectors~\cite{bennett-pra-1996}.  The number of
independent parameters $\lambda_j$ can be reduced to three
by a global phase change.

This transformation matrix can be readily  transformed into
the product basis
$\{\vert 00\rangle, \vert 01\rangle, \vert 10\rangle, 
\vert 11\rangle\}$ to obtain 
\renewcommand{\arraystretch}{1.5}
\begin{equation}
U_d^{\rm Prod}\!=\!\left(\!\begin{array}{cccc}
\cos{\frac{\alpha+\gamma}{2}}&0&0&\!\!i\sin{\frac{\alpha+\gamma}{2}}\cr
0&\!\!\!\!\!\!\cos{\frac{\alpha-\gamma}{2}}e^{-i\beta}&
i \sin{\frac{\alpha-\gamma}{2}} e^{-i\beta}&0\cr
0&\!\!\!\!\!\!i\sin{\frac{\alpha-\gamma}{2}}e^{-i\beta}& 
\cos{\frac{\alpha-\gamma}{2}}e^{-i\beta}&0\cr
i \sin{\frac{\alpha+\gamma}{2}}&0&0&\!\!\cos{\frac{\alpha+\gamma}{2}}
\end{array}
\!\right)
\end{equation}
\renewcommand{\arraystretch}{1}
where 
\begin{eqnarray}
\alpha&=&\phantom{-}(\lambda_1-\lambda_2-\lambda_3+\lambda_4)/2, \nonumber \\
\beta&=&-(\lambda_1+\lambda_2+\lambda_3+\lambda_4)/2, \nonumber \\
\gamma&=&\phantom{-}(\lambda_1-\lambda_2+\lambda_3-\lambda_4)/2
\end{eqnarray}

We are considering a quantum channel for a single qubit with another
qubit acting as the environment and a  general unitary
transformation (described above) providing the interaction
between the two qubits.  It is easy to see that the
properties of the channel do not change with the local
unitary transformations $U_1, V_1$ and $U_2, V_2$.
Therefore, for the analysis of the family of such channels,
the interaction unitary that needs to be considered is the
three-parameter family given in~(\ref{ch9_t}). 
The final output state of the system $\rho_{\rm out}$ emerging
from the channel after the action of $U_d$ is obtained by
tracing over the environment qubit.

Mathematically, a one-qubit channel can be described in a
number of ways, and we find it useful to picture it in terms
of an affine transformation of the Bloch sphere.
The quantum states of a single qubit on the Bloch sphere
can be represented as follows: 
\begin{equation} \rho =
\frac{1}{2}
(I+
{\mathbf n.\sigma})
\label{ch1_t} 
\end{equation} 
where the
$\sigma$'s are Pauli matrices, and ${\mathbf n}$ is a 3-component
real vector.  For pure states $\vert {\mathbf n} \vert=1$, and for 
mixed states $\vert {\mathbf n} \vert <1$.  Any arbitrary trace-preserving
quantum operation on~(\ref{ch1_t}) is given by a map of the form
$\rho\rightarrow \rho^\prime$, with  the new vector determining
$\rho^{\prime}$ given by 
\begin{equation} n_i^\prime=\sum^3_{j=1}
M_{ij} n_j + C_i.  \label{ch2_t} 
\end{equation} 
where $M_{ij}$ are the nine components of a $3 \times 3$
real matrix $M$ and $C_i$ are the three components of a
constant real column vector $C$.  This map, called the 
affine map, maps the Bloch sphere
(including its interior) onto a shifted ellipsoid with its
major axis less than
$2$~\cite{nielson-book-2002,jordan-pra-2005},

The total number of independent
parameters defining this map are twelve.  The matrix $M$
amounts to  a combination of proper rotations and
contractions in different directions of the vectors on the
Bloch sphere. The  vector $C$ corresponds to a shift in the
origin of the Bloch sphere.  

Two channels which differ from
each other by a unitary transformation of the qubit before
and after the action of the channel are identical.
Thus, we can utilize this freedom of 
performing arbitrary proper
rotations of the Bloch sphere before and after the channel
action to simplify the affine transformations for
the channel. This converts  the matrix
$M$ into a `singular-value' form with singular values
appearing along the diagonal.  
\begin{equation} M_d = R\, M\, S
\label{ch4_t} 
\end{equation} 
where $R$ and $S$ are real
orthogonal matrices with unit determinant and $M_d$ is a diagonal
matrix with the squares of its three diagonal elements given by  the 
eigen values of the positive semi-definite matrix
$M^{\dagger}M$.  
The shift vector $C$ is mapped onto another such
vector under these two proper rotations. After
utilizing this freedom through the action of $R$
and $S$,  the number of independent parameters for
single qubit channels is clearly six: namely the
three singular values of the matrix $M_d$, plus
the three components of the vector $C$.  There is
a further restriction of complete positivity on
the allowed affine transformations; however, that
does not reduce the number of independent
parameters~\cite{nielson-book-2002,bob-pra-1998,choi-alg-1975}.
These affine transformations  provide us with 
a complete description of one-qubit channels.

We now turn to an explicit calculation to obtain
the affine transformation given in
equation~(\ref{ch2_t}) for the class of channels
simulatable by a one-qubit mixed state environment
described in Figure~(\ref{channel}) (taking a
general initial system state $\rho=\vert
\psi\rangle \langle \psi\vert$ to $\rho_{\rm
out}$). For this purpose, we consider six points
on the input Bloch sphere corresponding to the
eigen states of the operators $S_x,S_y$ and $S_z$
and calculate the output density matrix 
for each of
these points.
At the end of this lengthy algebraic calculation,
the affine transformation turns out to be
\begin{equation}
M=\left(\begin{array}{lcr}
\cos{\alpha}\cos{\beta}&
-\lambda\cos{\alpha}&
-\lambda
\sin{\alpha}
\cos{\beta}
\cr
&
\sin{\beta}
\cos{\xi}
&
\sin{\eta}
\sin{\xi}
\cr\cr
\lambda\sin{\beta}
&
\cos{\beta}\cos{\gamma}
&
-\lambda\cos{\beta}
\sin{\gamma}
\cr
\cos{\gamma}
\cos{\xi}
&
&
\cos{\eta}
\sin{\xi}
\cr
\cr
\lambda
\sin{\alpha}
\cos{\gamma}
&
\lambda
\cos{\alpha}
\sin{\gamma}
&
\cr
\sin{\xi}
\sin{\eta}
&
\sin{\xi} 
\cos{\eta} 
&
\cos{\gamma}\cos{\alpha}
\end{array}
\right)
\label{ch17_t}
\end{equation}
The corresponding shift is
\renewcommand{\arraystretch}{1.2}
\begin{equation}
C=\left(
\begin{array}{l}
-\lambda\sin{\alpha}\sin{\beta}\sin{\xi}\cos{\eta}\cr
\phantom{-}\lambda\sin{\beta}\sin{\gamma}\sin{\xi}\sin{\eta}\cr
-\lambda\sin{\gamma}\sin{\alpha}\cos{\xi}
\end{array}\right)
\label{ch18_t}
\end{equation}
\renewcommand{\arraystretch}{1}
This affine transformation gives a
parameterization of all the channels simulated by
a one-qubit mixed state environment. We have thus
obtained a closed form expression for the complete
class of channels for a single qubit simulated by a
one-qubit environment.  This class of channels is
a six-parameter family with the six parameters
being $\alpha,\beta,\gamma,\eta,\xi$ and $\lambda$.  We obtained
this form by computing the action of the channel
explicitly on several test states until the
transformation is uniquely determined.

As we have observed,  the number of independent
parameters  for a single qubit channel simulated
using this mixed state environment also turns out
to be six.  It had thus been conjectured that all
single qubit channels may be simulatable using a
single qubit mixed state as the
environment~\cite{lloyd-science-1996}. Despite
counter-examples like the two-Pauli channel, the
possibility that the set of single qubit channels
simulated using a mixed state environment may
occupy a sizable volume in the space of single
qubit channels remains open, and it may turn out
that the counter examples are a set of measure
zero in the channel space. This is precisely the
question that will be explored and resolved in the
following section, using
the closed form expression given in
equations(~\ref{ch17_t}) and~(\ref{ch18_t}).
\section{The generalized depolarizing channel}
\label{depolarizing-channel}
The general expression for the family of channels obtained in the
previous section is very useful, and  several special cases are of
particular interest. We consider the case of generalized
depolarizing channels  defined as the ones for which the origin
of the Bloch sphere is not shifted. This class is simple and we
know that it contains non-trivial examples which cannot be
implemented using a one-qubit pure state environment. We therefore
analyze this sub-class in some detail. The 
operator sum representation of this class is given by
\begin{equation}
\rho\rightarrow
\rho^{\prime}=\sum^{3}_{j=0}\epsilon_j A_j \rho
A_j^{\dagger}
\label{dp-kraus}
\end{equation}
Where $A_0$ is the identity matrix and
for $j=1,2,3$ we have
$A_j=\sigma_j$ and we have
$\epsilon_0+\epsilon_1+\epsilon_2+\epsilon_3=1$.
The corresponding affine transformation is
\begin{eqnarray}
\!M^{\rm DP}\!=\!\left(
\begin{array}{ccc}
\!\epsilon_{0}\!+\!\epsilon_{1}\!-\!\epsilon_{2}\!-\!\epsilon_{3}
& \!\!\!\!\!\!\!0&\!\!\!\!\!\!\!0\\
\!0&\!\!\!\!\!\!\!\epsilon_{0}\!-\!\epsilon_{1}\!+
\!\epsilon_{2}\!-\!\epsilon_{3}
&\!\!\!\!\!\!\!0\\
\!0&\!\!\!\!\!\!\!0&
\!\!\!\!\!\!\!\epsilon_{0}\!-\!\epsilon_{1}\!-\!\epsilon_{2}+\epsilon_{3}
\end{array} 
\right)
\nonumber
\nonumber \\
\label{dp-affine}
\end{eqnarray}
The shift $C$ is zero indicating that it is
indeed a generalized depolarizing channel.  The
three diagonal entries of the matrix $M^{\rm DP}$ take
values such that this family of generalized
depolarizing channels is geometrically represented
by a tetrahedron volume with vertices given by
$(1,-1,-1)$ $(-1,1,-1)$ $(-1,-1,1)$ $(1,1,1)$.
Each value of $x,y,z$ which lies inside this
volume is an allowed set of diagonal elements for
the affine transformation $M^{DP}$ for the
generalized depolarizing 
channel~\cite{terhal-pra-1999,
bob-pra-1998,bourdon-pra-2004}. The
volume of this tetrahedron is $\frac{8}{3}$.  
The vertices of the tetrahedron represent a
unitary map for which we require only one operator in
the operator sum representation. The edges represent the
two operator maps and the faces represent three operators
maps while the points inside the tetrahedron 
require all the four operators for their realization. 
The families of two-Pauli channels which will be 
discussed in the next sub-section  are represented
on the faces of the tetrahedron (not including the
edges).

The affine transformation
corresponding to one-qubit generalized
depolarizing channels,
with a single qubit as the environment, can be obtained
by setting the shift vector $C$
to zero and thereby simplifying the 
equation~(\ref{ch18_t}).  
It turns out that this
can be achieved in eleven different ways, leading to eleven
different cases. For each case, we compute the singular
values of the affine transformation which suffice to
characterize the channel. Although the algebra is a little
involved, the final result in each case turns out to be
rather
simple. In each case the affine transformation 
can be brought to the following form, by local
unitaries before and after the channel action:
\begin{equation}
M^{\rm DP}_{\rm One-qubit}=
\left(\begin{array}{ccc} \cos a \cos b& 0 & 0\\
0&\cos b \cos c&0\\0&0&
\cos c\cos a \end{array} \right)
\label{gdpol-affine}
\end{equation}
where $a,b$ $c$ depend upon the six parameters of
the channel namely $\alpha,\beta,\gamma,\xi,\eta$, and
$\lambda$.

One can immediately see that the normal depolarizing
channel  which maps the entire Bloch sphere
onto the origin,  and
which cannot be simulated  using a one-qubit pure state
environment, is contained here. It corresponds to
any pair from $a,b,c$ taking values $\pi/2$.
However, the important question we would like to
address is how many generalized depolarizing
channels are contained in the affine
transformation~(\ref{gdpol-affine}).
\subsection{Volume issues}
We are interested in finding the volume inside the
tetrahedron occupied by the affine transformation
represented by equation~(\ref{gdpol-affine}).
This will help us explore the fraction of
depolarizing channels simulatable using a
one-qubit mixed state environment.

\begin{figure}
\includegraphics[scale=0.55,angle=-90]{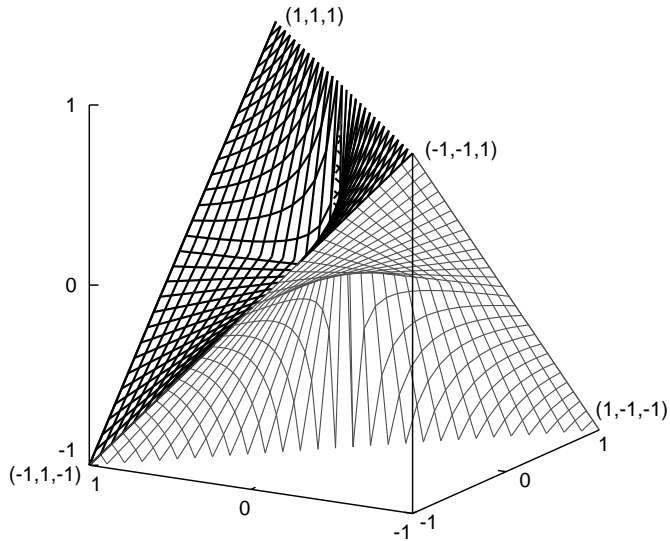}
\caption{ \label{channel-volume-1} Graphical
representation of the volume occupied by channels
simulatable by a one-qubit mixed state environment.
This volume is $\frac{3}{8}$ of the volume of the
tetrahedron which represents all possible
generalized depolarizing channels. All the
vertices and edges of the tetrahedron are
contained while no other point on any of the
faces is contained.}
\end{figure}

\begin{figure}
\includegraphics[scale=1,angle=0]{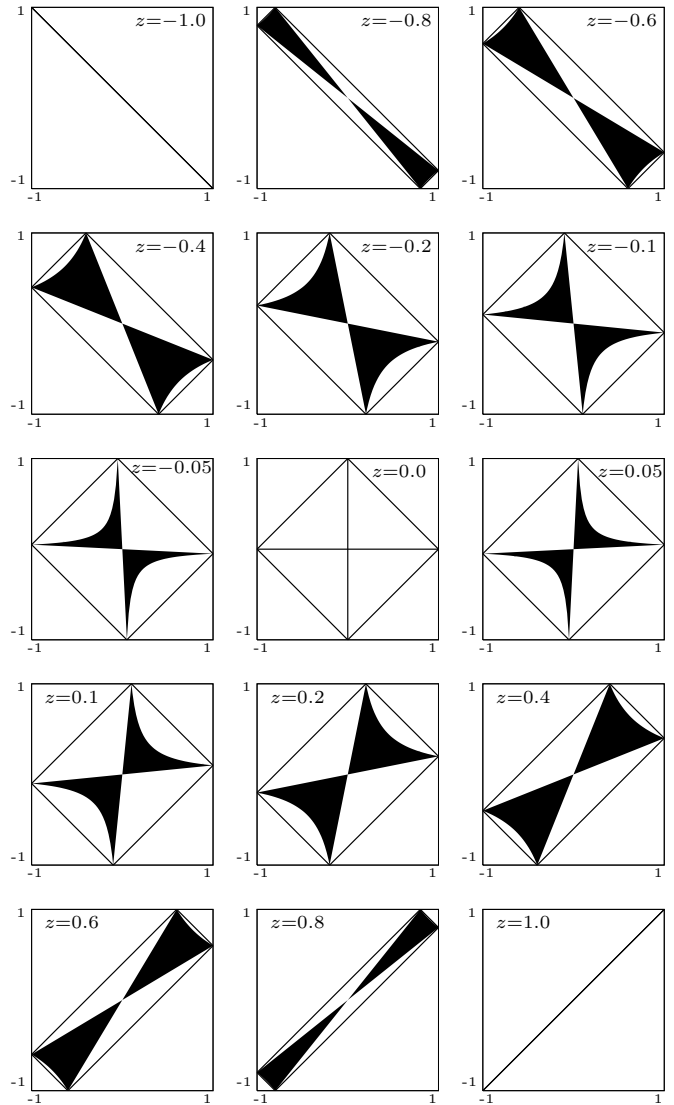}
\caption{ \label{channel-volume-2} Cross sections
of the volume depicted in
Figure~\ref{channel-volume-1} for different values
of $z$ from $-1$ to $+1$. For each value of $z$
the rectangle is the cross section of the
tetrahedron and the shaded area is the fraction
simulatable with a one-qubit mixed state
environment. The shaded fraction approaches zero
as $\vert z \vert$ approaches $0$ and it approaches
one as  $\vert z \vert$ approaches $1$. In
between for a given $z=z_0$  the darkened area  is
bounded by hyperbola $xy=z_0$ and lines $y= z_0 x$
and $y= x/z_0$.} 
\end{figure}

The constraints which are imposed by the structure
of equation~(\ref{gdpol-affine}) can be
analytically worked out and turn out to be
\begin{equation} 
x y \leq z;\quad  y z \leq x;
\quad  z x \leq y  
\end{equation} 
where $x=\cos{a}\cos{b}, y=\cos{b}\cos{c}$ and 
$z=\cos{c}\cos{a}$.
The volume
constrained by these conditions is shown in
Figure~(\ref{channel-volume-1}). All the vertices
and the edges of the tetrahedron are touched.
However no point on any of the faces is
contained. It can be visualized as if each face 
of the tetrahedron has been scooped out and
the depth of this scoop extends all the way to the
centroid. The volume enclosed by this shape
can be computed and it turns out to
be $1$. 
This implies that only $\frac{3}{8}$ 
of the generalized depolarizing channels can be
simulated by a one-qubit mixed state environment.
To give a more visual feel for this volume we depict
its cross sections for different $z$ values in
Figure~(\ref{channel-volume-2}). At each value of
$z$ the tetrahedron cross section is a rectangle
and the darkened area is the cross section of the
volume depicted in
Figure~(\ref{channel-volume-1}) (cross sectional
views for constant $x$ and $y$ will look 
identical).
\subsection{Two-Pauli channel}
We now turn to an interesting sub-family of
depolarizing channels, namely, the two-Pauli
channel.  Two-pauli channels are the ones for
which only one of $\{\epsilon_1,\epsilon_2,
\epsilon_3\}$ is zero and they are  represented on
the faces of the tetrahedron (leaving out the edges).

Further, for simplicity we restrict ourselves to 
a special case of a one-parameter sub-family of
two-Pauli channels for which the nonzero elements
are given by
$\epsilon_2=0$ and 
$\epsilon_1=\epsilon_2= (1-\kappa)/2$ and
$\epsilon_0=\kappa$.
with $0\leq \kappa \leq 1$. 
The affine transformation for 
the two-Pauli channel can be easily computed and is 
given by 
\begin{equation}
M^{\rm 2-Pauli}=\left(\begin{array}{lcr}
\kappa\phantom{-1}&0&0\cr
0&\kappa&0\cr
0&0&2\kappa-1 
\end{array}
\right)
\label{two-pauli-affine}
\end{equation}
This is a special case of the generalized
depolarizing channels which is represented by a
line on the face of the tetrahedron joining the
vertex $(1,1,1)$ to the mid point of the edge
below it. We take this particular case because it
has been discussed in detail in the literature.

We will directly argue that the affine
transformation~(\ref{two-pauli-affine}) is not a
special case of the affine
transformation~(\ref{gdpol-affine}) for some
values of $a$, $b$ and $c$. For the
matrix~(\ref{gdpol-affine}) to take the
form~(\ref{two-pauli-affine}), two of its entries
should become equal; let us assume that first two
are equal. This forces $\cos a =\cos c$ and this
further implies that $cos a$ is a factor of all
the singular values. However, the affine
transformation~(\ref{two-pauli-affine}) does not
have a non-trivial factor and hence the only
possibility is $\cos a=1$. Now we impose the
relationship between the equal and unequal
singular values leading to $2 \cos b -1 =1$ which
implies that $\cos b=1 $. A similar result is
obtained if we begin by equating any other pair of
singular values in equation~(\ref{gdpol-affine}).
The above analysis shows that the only place where
the two-Pauli affine transformation can match the
affine transformation~(\ref{gdpol-affine}) is when
$\kappa=1$, at which point the two-Pauli channel
disappears and one is left with the identity
transformation.  Thus we conclude that the
two-Pauli channel is not a special case
of~(\ref{gdpol-affine}). This result has been
obtained earlier, where it was shown that the
two-Pauli channel is a counter-example to the
conjecture that a one-qubit mixed state
environment can simulate all one-qubit quantum
channels. Our demonstration is simpler
and analytical, since we have obtained the explicit
expression for the affine transformation. 
It is not difficult to generalize this argument
for the other two-Pauli channels.  Results about
the points on the face of the tetrahedron not
modelable using a one-qubit mixed state environment,
are actually contained in the volume analysis
of the previous section.  We have presented the
case of two-Pauli channels in detail due to the
simplicity of the argument and the fact that
two-Pauli channels have been discussed in the
literature~\cite{terhal-pra-1999,bob-pra-2006}.
\section{Concluding Remarks}
\label{conclusions}
We have obtained a canonical form corresponding to
single qubit channels with a one-qubit mixed state
modeling the environment.  This parameterization
has been obtained analytically by computing the
affine transformation corresponding to this class
of channels. Applying our results to the sub-class
of depolarizing channels, we found that
although a sizable
volume ($\frac{3}{8}$ of the total volume)
is modeled by a one-qubit mixed state environment,
there is a finite volume in the channel space
($\frac{5}{8}$ of the total volume)
which cannot be modeled in this fashion.  The special
case of the two-Pauli channel discussed in the
literature has been shown to be in this missing
volume and hence is not simulatable via a one-qubit mixed state
environment.  The results were achieved by starting with
the general expression for the affine
transformation, restricting it to the case with
zero shift, and computing the singular values. It
turns out that we can arrive at the precise
constraints on the volume analytically, thereby
producing a picture of the subset of
generalized depolarizing channels simulatable by
a one-qubit mixed state in the environment.

The affine transformation which we obtain is
general and one can also explore channels other
than the depolarizing channels.  The analysis
of the volume occupied by such channels in the
entire channel space for one qubit is an involved
problem and will be taken up elsewhere. We note
that a similar conclusion about the two-Pauli
channel from a very different perspective has
been obtained in~\cite{bob-pra-2006}. 

\begin{acknowledgments} Geetu Narang
thanks  R. B. Griffiths for discussions and providing financial
support and hospitality at Carnegie Mellon
University.DST India is acknowledged for
financial support.
\end{acknowledgments}

\end{document}